\documentclass[journal]{IEEEtran}

\usepackage[pdftex]{graphicx} 
\graphicspath{{fig/}}
\DeclareGraphicsExtensions{.pdf,.png,.jpg}

\usepackage{import} 
\usepackage{color}
\usepackage{transparent} 

\usepackage[cmex10]{amsmath} 

\usepackage{cite} 

\usepackage{array} 

\usepackage{fixltx2e}

\usepackage{verbatim}  

\begin{document}
\title{GeFRO: a New Charge Sensitive Amplifier Design with a Minimal Number of Front-end Components}

\author{L.~Cassina, C.~Cattadori, A.~Giachero, C.~Gotti, M.~Maino, G.~Pessina
\thanks{L.~Cassina, C.~Cattadori, A.~Giachero, C.~Gotti, M.~Maino and G.~Pessina are with the INFN Milano Bicocca and Department of Physics, University of Milano Bicocca, Piazza della Scienza 3, Milano, 20126, Italy.}
\thanks{E-mail: claudio.gotti@mib.infn.it}
}

\maketitle
\pagestyle{empty}
\thispagestyle{empty}

\begin{abstract}
A new approach was developed for the design of front-end circuits for semiconductor radiation detectors.
The readout scheme consists of a first stage made of only a few components located close to the detector, and of a remote second stage located far from the detector, several meters away.
The second stage amplifies the signals from the first stage and closes the feedback loop to discharge the input node after each event.
The circuit has two outputs: one gives a ``fast'' signal, with a bandwidth larger than \mbox{20 MHz}, allowing to preserve the high frequency components of the detector signals, which may be useful for timing measurements, pile-up rejection or pulse shape discrimination.
The second output gives a ``slow'' signal, whose gain depends only on the value of the feedback capacitor, as happens with a classic charge sensitive amplifier, allowing to obtain higher resolution and lower drift.
The prototype was named GeFRO for Germanium front-end, and was tested with a BEGe detector from Canberra.
The wide bandwidth of the ``fast'' signal gave a timing resolution of the order of \mbox{20 ns}.
The noise of the circuit at the ``slow'' output after a \mbox{10 $\mu$s} Gaussian shaping was close to \mbox{160 e$^{-}$ RMS} with an input capacitance of \mbox{26 pF}.

\end{abstract}

\section{Introduction}
\IEEEPARstart{S}{emiconductor} detectors are used since a long time to detect ionizing radiation.
Among these, Germanium detectors are extensively used for gamma spectroscopy, which requires full absorption of incoming radiation and high energy resolution.
Other semiconductor materials exist which can be used for the same purpose.
As well known, such detectors are generally readout with charge amplifiers, which convert the charge signals from the detector to voltage signals whose amplitude depends on the value of a known feedback capacitor \cite{chargeamp1, chargeamp2, chargeamp3}.
A charge amplifier for the readout of semiconductor detectors is generally made of an input JFET (or MOSFET) selected for low noise, a second amplification stage, and the parallel combination of a capacitor and a large value resistor as the feedback elements.
For the best noise performance the capacitance at the input node must be minimized, and thus the input JFET and the feedback components should ideally be placed very close to the detector.

For some applications the classic configuration may present some drawbacks.
If the detector is operated in a cryogenic environment (as happens with Germanium detectors) and the circuit is placed close to the detector, then the circuit must operate at low temperature.
While this is not generally a problem for the input JFET and the feedback components, provided that they are selected accordingly, it can complicate the design of the second stage, possibly preventing the use of some operational amplifiers.
If several detectors are placed close to each other and the readout circuits have wide bandwidth, the power dissipation in the cryostat may be large.
Moreover, large power supply bypass capacitors may be required at cold to prevent disturbances or crosstalk from being injected through the supply voltages, especially in the case where the connecting cables are long and have a non negligible series resistance.
High levels of radiation near the detectors may also require the electronic components to be radiation tolerant, and may prevent the operators from accessing the circuits for optimization or maintenance.
In some cases it can thus be beneficial to separate the first stage, made of the JFET and the feedback components, from the second stage, with the aim of placing it in a less hostile environment.
In the case of the classic charge sensitive amplifier this results in a bandwidth upper limit, since the cables connecting the first and second stage add capacitance to the feedback loop.
Even if the effect of cable capacitance is minimized by connecting the first and second stage with properly terminated transmission lines, the rise time of the signals cannot be shorter than a few times the propagation delay along the cables, as will be discussed in the following.

If the first and second stage of the charge amplifier are separated by a large distance but the high frequency signal components must be retained, a dual readout approach can be used.
The ``slow'' output of a classic charge amplifier (anyway optimized for the maximum achievable closed-loop bandwidth) can be used to measure the total deposited charge, while a ``fast'' open loop output provides the complementary information at high frequency.
The core ideas behind the ``fast'' output were already described in previous publications \cite{gefro1, gefro2}.
In this paper several circuit improvements were introduced to overcome open issues related to small variations of the capacitance at the input node, which could be caused by variations in the counting rate or by mechanical vibrations.
The improved circuit design is explained in this paper in thorough detail, and the results obtained with a prototype coupled to a small anode Germanium detector, a Canberra BEGe, are presented.

The field which may benefit most from this circuit solution is that of rare event searches with Germanium detectors, and particularly neutrinoless double beta decay search experiments such as GERDA \cite{gerda1} and MAJORANA \cite{majorana1}.
The readout approach described in this paper was in fact developed in the framework of the R\&D for the phase II of the GERDA experiment, and is then particularly tailored for such application.
These experiments need extremely low background coming from the environment, including the detectors themselves and their readout electronics, in order to achieve the necessary sensitivity to the rare nuclear decays of interest.
Particular care must be taken in realizing a radiopure readout chain, and a means to this is to minimize the number of components close to the detectors by placing the second stage outside the cryostat, several meters away, while keeping the input JFET and the feedback components close to the detectors to satisfy the low noise requirements.
At the same time, the small anode Germanium detectors which these experiments plan to deploy allow to discriminate between different types of particle interactions by pulse shape analysis on the rise time of the signals, reducing the background in the energy region of interest \cite{psa1, psa2, psa3}.
The readout circuit should then provide the necessary bandwidth of a few tens of MHz in order to preserve the charge collection profile in the detectors.
Anyway, as already mentioned, the design approach described in this paper can be applied in other fields, whenever a no-compromise solution with a minimal number of front-end components close to the detector may be of interest.

\section{The GeFRO Circuit}

\begin{figure*}
\centering
\def\svgwidth{\linewidth}
\fontfamily{ppt}\selectfont \scriptsize
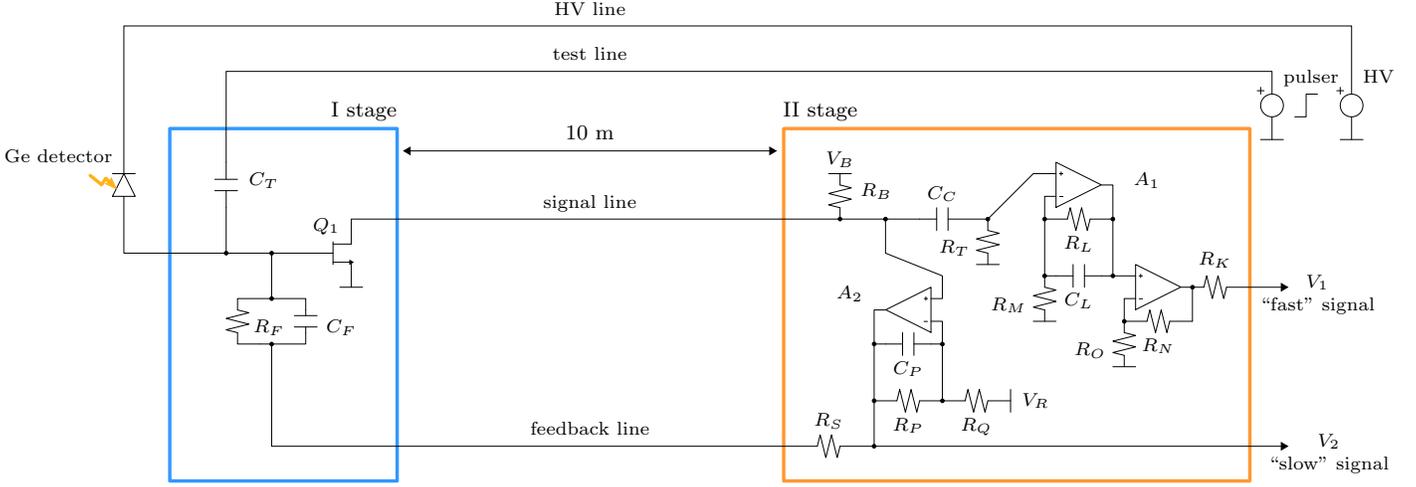
\caption{Basic schematic of the GeFRO circuit.}
\label{fig:GeFROSchematic}
\end{figure*}

The schematic of the proposed circuit solution is shown in figure \ref{fig:GeFROSchematic}.
It consists of a small front-end stage located close to the detector and a remote second stage located several meters away and connected to the first stage through terminated transmission lines.
As already mentioned, the circuit was proposed for the phase II of GERDA, which employs Germanium detectors operated in a cryogenic environment with low radioactive background, to which the readout electronics must comply.
Due to the dimensions of the cryostat, the distance between the detectors and the room where the second stage electronics could be placed is about ten meters.

The front-end stage is composed of a JFET $Q_1$ and the feedback elements $R_F$ and $C_F$.
The input JFET is operated in common source configuration and is located close to the detector to minimize the stray input capacitance, so it should be chosen to have low noise at cryogenic temperature.
Several devices are commercially available for such purpose from many manufacturers.
Due to the radiopurity constraints of our application, our choice was narrowed to the devices which could be purchased in bare Silicon die.
At a few mA of bias current, most of the JFETs tested at low temperature have a transconductance of about \mbox{2 mA/V} per pF of input capacitance.
In order to have a transconductance larger than \mbox{20 mA/V}, necessary to drive the signal on a terminated \mbox{50 $\Omega$} transmission line without significant gain loss, the resulting input capacitance from $Q_1$ is at least \mbox{10 pF}.
In the case where the detector capacitance is much smaller, it would then be convenient to work in mismatch conditions, with the input capacitance dominated by $Q_1$.
In the measurements presented in the following sections a SF291 JFET from Semefab was used as $Q_1$, featuring a transconductance \mbox{$g_m =$ 33 mA/V} at \mbox{$I_D=$ 10 mA} and \mbox{$V_{DS}=$ 2 V}, measured at \mbox{77 K} with a Keithley 4200 semiconductor analyzer.
Its gate capacitance of \mbox{16 pF} was instead evaluated from the amplitude of the ``fast'' GeFRO signal, as will be described in section \ref{sec:SigOsc}.
In the measurements with the BEGe detector presented in this paper, the total input capacitance $C_I$ was close to \mbox{26 pF}, mainly contributed by the input JFET and parasitics.

At DC, assuming $R_P \gg R_Q$, the drain of $Q_1$ is held at $V_R$ by the feedback loop through $A_2$, and its bias current is given by $\left( V_B-V_R\right) / R_B$.
In our case \mbox{$V_B=$ 12 V}, \mbox{$V_R=$ 2 V}, then \mbox{$R_B=$ 1 k$\Omega$} gives for $Q_1$ a bias current of \mbox{10 mA}.
Notably, the bias voltage and current of the input transistor can be tuned by acting on the second stage, without direct access to the front-end.
The DC voltage on the feedback line is $V_G - i_L R_F$, where $V_G$ is the gate voltage of the polarized JFET and $i_L$ is the total current flowing through $R_F$, given by the sum of the leakage currents from the detector and from the JFET.
In our measurements $i_L$ was contributed by the current in the detector, a few pA at low event rates, and by the gate current of the JFET at \mbox{77 K}, much less than \mbox{1 pA}.

Concerning the feedback components, the values chosen for our prototype are \mbox{$R_F=$ 500 M$\Omega$}, \mbox{$C_F=$ 0.9 pF} ($C_F$ already includes the parasitic contributions due to $R_F$ and layout).
In the first version of the GeFRO, as reported in \cite{gefro1, gefro2}, only a Schottky diode was used as the feedback element, which served at the same time as the (non linear) large value resistor and capacitor.
It was then found difficult to find a Schottky diode with a high degree of radiopurity, while on the contrary Silicon resistors in bare die of high radiopurity were found to be commercially available, with values ranging up to \mbox{150 M$\Omega$} (MSTF 6SS-15005 J-E from Mini-Systems, Inc.).
Three of such resistors can be used in series to obtain a feedback resistor of the required value.
If a resistor is used for $R_F$ then the Schottky diode is no longer necessary.

A normal Silicon diode of low capacitance, which can be easily found in bare die, can be used at the input to protect the JFET against negative overvoltages which may occur if the high voltage bias of the detector is decreased too fast.
If \mbox{$V_G<$ 0 V} the diode needs a negative reference voltage, which would require an additional cable to the front-end stage.
As an alternative, the protection diode can be connected in parallel with $R_F$ (and biased with \mbox{0 V} in this case), but in that case the capacitance of the diode should be considered in parallel with $C_F$ in the following evaluations.
Anyway, if the capacitance of the detector is negligible with respect to the input capacitance due to the JFET and parasitics, as happens with small anode detectors such as the Canberra BEGe, then accidental negative overvoltages at the input are strongly attenuated, and the diode can be omitted.
For this reason it will not be considered in the rest of this paper.

The schematic of figure \ref{fig:GeFROSchematic} also shows the line used to inject a test signal through the test capacitance $C_T$ located close to the input node, whose value was \mbox{0.5 pF}.
If the test signal is a voltage step of amplitude $V_T$, a current pulse carrying a known charge $Q = C_T V_T$ is generated.

A signal from the detector is a current pulse carrying a given amount of charge $Q$.
The feedback loop has a finite bandwidth, and has no effect on the first part of the signals.
On the fast transient, the current signal is integrated on the input capacitance $C_I$ and becomes a voltage step at the input of $Q_1$.
The JFET converts this voltage step at the input to a current step at its output, which is driven into a terminated signal line of characteristic impedance $R_T$ to avoid reflections.
If the drain of the JFET can be approximated as an ideal current source (which is true if its drain-source resistance and gate-drain capacitance can be neglected) then the output signal is insensitive to the series resistance of the output line, which can be as high as a few $\Omega$/m if coaxial cables of small section are used.
The line termination $R_T$ is AC coupled through $C_C$.
Its value is expected to be \mbox{50-100 $\Omega$}, depending on the choice of cables.
In the rest of the paper a characteristic impedance of \mbox{50 $\Omega$} will be considered.
The signal across $R_T$ is amplified by the fast voltage amplifier $A_1$ and constitutes the ``fast'' signal, expressed by
\begin{equation}
V_1 (t) = - G_1 \frac{Q}{C_I} g_m R_T \textrm{e}^{ -\frac{t}{\tau_L-\tau_D}}
\label{eq:V1(t)1}
\end{equation}
where $G_1$ is the gain of $A_1$, $g_m$ is the transconductance of $Q_1$, and $\tau_L-\tau_D$ is the time constant related to the discharge of the input node after each event, which gives the fall time constant of the ``fast'' signal.
The value of $\tau_L$ is determined by the bandwidth of the feedback loop, while the value of $\tau_D$ is determined by the length of the cables connecting the first and second stage, as will be calculated in the following sections.
For stability constraints it will be shown that $\tau_L$ must always be a few times larger than $\tau_D$.
The rise time of the ``fast'' signal is determined by the bandwidth of the signal cables and of $A_1$.
In our prototype $A_1$ was based on a LT6230-10 operational amplifier from Linear, chosen for low voltage noise and large bandwidth, operated at a gain of \mbox{21 V/V} (\mbox{$R_L=$ 1 k$\Omega$}, \mbox{$R_M=$ 50 $\Omega$}) and externally compensated with \mbox{$C_L=$ 10 pF} to obtain a bandwidth of about \mbox{20 MHz}.
The output of the LT6230-10 was amplified by a AD811 operational amplifier from Analog Devices with a gain of \mbox{6.4 V/V} (\mbox{$R_N=$ 270 $\Omega$}, \mbox{$R_O=$ 50 $\Omega$}), so that the ``fast'' signal at the output of the second stage could be driven over a terminated \mbox{50 $\Omega$} transmission line (\mbox{$R_K=$ 50 $\Omega$}) with an overall gain \mbox{$G_1=$ 67 V/V}.

On a longer time scale the feedback loop becomes effective, forcing the discharge of the input node through the feedback components $C_F$ and $R_F$.
To do this the feedback amplifier $A_2$ injects a charge through $C_F$ which counterbalances the input charge $Q$.
When the input node is discharged, a voltage $-Q / C_F$ is found across $C_F$, which then discharges through $R_F$ with time constant $C_F R_F$.
The gain and bandwidth of $A_2$ are determined by $C_P$, $R_P$ and $R_Q$.
Acting on their values allows to tune $\tau_L$ to assure the stability of the feedback loop.
The fall time of the ``fast'' signal $V_1$ coincides with the rise time of the feedback signal $V_2$, which constitutes the ``slow'' signal.
The feedback amplifier in our prototype was a AD797 operational amplifier from Analog Devices, chosen for low voltage noise, low DC offset and relatively large bandwidth.
The resistor $R_S$ is used to terminate the feedback line on its characteristic impedance at one end, avoiding reflections which may impact stability.
If the signal and feedback cables have the same characteristic impedance, then $R_S = R_T$.
Neglecting the DC voltage on the feedback line, the ``slow'' signal in response to a charge $Q$ is given by
\begin{equation}
V_2 (t) = - \frac{Q}{C_F} \frac{\tau_F}{\tau_F - \tau_L + \tau_D} \left( \textrm{e}^{-\frac{t}{\tau_F}} - \textrm{e}^{- \frac{t}{\tau_L-\tau_D}}  \right)
\label{eq:V2(t)1}
\end{equation}
where $\tau_F = C_F R_F$, that is the signal of a classic charge amplifier with bandwidth limited to $1/2 \pi (\tau_L-\tau_D)$.
It should be noted that the gain of the ``slow'' signal depends mainly on the value of $C_F$.
The input capacitance $C_I$ affects the value of $\tau_L$, as will be shown, but the effect on the signal amplitude is a second order contribution if \mbox{$\tau_F \gg \tau_L$} and the shaping is slow enough with respect to $\tau_L$.
So, even if its bandwidth is smaller with respect to the ``fast'' signal, the ``slow'' signal can give a more precise and reliable estimate of the input charge $Q$ in the case where the capacitance $C_I$ fluctuates.

\begin{figure}
\centering
\def\svgwidth{\linewidth}
\fontfamily{ppt}\selectfont \scriptsize
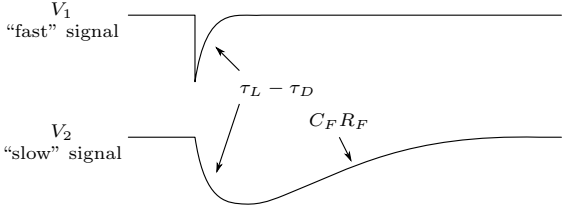
\caption{``Fast'' and ``slow'' signals expected at the output of the GeFRO circuit.}
\label{fig:GeFROExpectedSignals}
\end{figure}

The ``fast'' and ``slow'' signals are schematically depicted in figure \ref{fig:GeFROExpectedSignals} for input pulses carrying a positive charge $Q$.
In a general application, both waveforms can be acquired at the same time.
The ``fast'' signal can be used to resolve the charge collection profile in the detector, while the ``slow'' signal can be used for energy measurement.
In the following sections the conditions for the stability of the loop gain will be considered and the validity of equations (\ref{eq:V1(t)1}) and (\ref{eq:V2(t)1}) will be demonstrated.

\section{Loop Gain and Stability}

Let us first consider the transfer function of the amplifier $A_2$ with its feedback components $C_P$, $R_P$ and $R_Q$.
In the domain of the complex frequency $s$ the transfer function of the open loop operational amplifier with a dominant pole can be modeled as
\begin{equation}
G^{OL}_2 (s) = \frac{A}{1+s \tau_2}
\label{eq:G2OL(s)1}
\end{equation}
where $A$ is the open loop gain at DC and $1/ 2 \pi \tau_2$ is the frequency of the dominant pole.
As well known, the gain-bandwidth product of the amplifier is $\omega_T / 2 \pi$, where \mbox{$\omega_T = A/ \tau_2$}.
The closed loop transfer function is then
\begin{equation}
G_2 (s) = \frac{R_P}{R_Q} \frac{1}{1+s R_P  \left( C_P + \frac{1}{\omega_T R_Q} \right)}
\label{eq:G2(s)1}
\end{equation}
Equation (\ref{eq:G2(s)1}) was obtained by approximating for $R_P \gg R_Q$ and is valid in the range of frequencies $f$ where
\begin{equation}
\frac{1}{2 \pi C_P R_Q} \gg f \gg \frac{1}{2 \pi \tau_2}
\label{eq:1/CPRQgg}
\end{equation}
Since the dominant pole of the operational amplifier is at very low frequency, the second inequality in (\ref{eq:1/CPRQgg}) is easily satisfied.
The first inequality in (\ref{eq:1/CPRQgg}) will instead be verified once the values for $C_P$ and $R_Q$ will be chosen.

Let us now consider the entire feedback loop between the first and second stage.
The loop gain $T(s)$ is given by
\begin{equation}
\begin{split}
T (s) = \left( \frac{1 + s C_F R_F}{1+s  C_I R_F} \right) \times \\
\times \left( - g_m R_B \frac{1+s C_C R_T}{1+s C_C R_B} \right) 
 G_2 (s)  \textrm{e}^{-s \tau_D}
\end{split}
\label{eq:T(s)1}
\end{equation}
The first term is due to $R_F$, which forms a pole with the total input capacitance to ground $C_I$ and a zero with the feedback capacitance $C_F$.
The expression was approximated for $C_I \gg C_F$, which is certainly true with the values given in the previous section.
The second term is due to the gain of the JFET on the total impedance it sees at its output.
This term contributes with a pole at $C_C R_B$ and a zero at $C_C R_T$.
This term was approximated for $R_B \gg R_T$, which is allowed if \mbox{$R_B=$ 1 k$\Omega$} and \mbox{$R_T=$ 50 $\Omega$} as in our case.
The third term is due to the feedback amplifier $A_2$, as calculated above.
The last exponential term represents the phase shift introduced by the propagation delay $\tau_D$ along the signal and feedback lines.
Assuming both lines to have length $L$, the propagation delay is given by
\begin{equation}
\tau_D = 2 L t_P
\end{equation}
where $t_P$ is the propagation delay per unit length.
If \mbox{$L=$ 10 m} and \mbox{$t_P=$ 5 ns/m} then \mbox{$\tau_D=$ 100 ns}.

\begin{figure}
\centering
\def\svgwidth{\linewidth}
\fontfamily{ppt}\selectfont \scriptsize
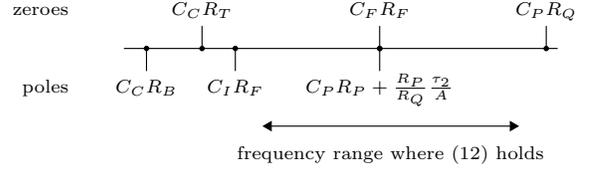
\caption{Poles and zeroes in the feedback loop of the GeFRO.}
\label{fig:GeFROPolesAndZeroes}
\end{figure}

The components whose values at this point are not fixed by other constraints are $C_C$, $R_P$, $C_P$ and $R_Q$.
Let us choose $C_C$ to be very large, say \mbox{1000 $\mu$F}.
With the values given above \mbox{$C_C R_B=$ 1 s}, \mbox{$C_C R_T=$ 50 ms}, \mbox{$C_I R_F=$ 13 ms},  \mbox{$C_F R_F=$ 450 $\mu$s}.
Above a few hundred Hz the loop gain can then be approximated as
\begin{equation}
T (s) = - \frac{R_P}{R_Q} \frac{g_m R_T}{sC_I R_F}
\frac{1 + s C_F R_F}{1+s R_P \left( C_P + \frac{1}{\omega_T R_Q} \right)}
  \textrm{e}^{-s \tau_D}
\label{eq:T(s)2}
\end{equation}
We can now choose $R_P$ so that
\begin{equation}
C_F R_F = R_P \left( C_P + \frac{1}{\omega_T R_Q} \right)
\label{eq:CFRF}
\end{equation}
With such choice equation (\ref{eq:T(s)2}) simplifies to
\begin{equation}
T (s) = -  \frac{R_P}{R_Q} \frac{g_m R_T}{sC_I R_F}  \textrm{e}^{-s \tau_D}
\label{eq:T(s)3}
\end{equation}
If we now define
\begin{equation}
\tau_L = \frac{R_Q}{R_P} \frac{C_I R_F}{g_m R_T}
\label{eq:tauL1}
\end{equation}
the loop gain (\ref{eq:T(s)3}) can then be written as
\begin{equation}
T (s) = - \frac{\textrm{e}^{-s \tau_D}}{s \tau_L}  
\label{eq:T(s)4}
\end{equation}
The loop gain clearly shows a dominant pole at low frequency, and a phase shift term related to the propagation delay along the transmission lines which connect the first and the second stage.
This simplified expression is valid in the range of frequencies shown in figure \ref{fig:GeFROPolesAndZeroes}.

As an alternative condition, the loop gain expressed by (\ref{eq:T(s)2}) can be approximated by letting $R_P \rightarrow \infty$.
In this case, at frequencies larger than $1/ 2 \pi C_F R_F$ (that is, above a few kHz) the loop gain can still be written as (\ref{eq:T(s)4}) where $\tau_L$ is now defined as
\begin{equation}
\tau_L = \frac{R_Q}{g_m R_T} \frac{C_I}{C_F} \left( C_P + \frac{1}{\omega_T R_Q} \right)
\label{eq:tauL2}
\end{equation}
In most of the measurements presented in this paper the first case will be preferred, that is $R_P$ satisfying (\ref{eq:CFRF}), but the following evaluations on the stability of the loop gain apply also to the case where $R_P \rightarrow \infty$.
As can be clearly seen, in both cases $\tau_L$ can be tuned by changing the values of $C_P$ and $R_Q$.

Let us now consider the stability of the loop gain.
At DC the loop gain as given by (\ref{eq:T(s)2}) is a negative real number, or in other words the phase of $-T(s)$ is 0$^{\circ}$.
The critical frequency to determine the stability of the feedback loop is that for which \mbox{$\left| T \right|=$ 1}, that is
\begin{equation}
f_{\left| T \right| = 1} = \frac{1}{2 \pi \tau_L}
\end{equation}
The closed loop transfer function is proportional to $1/\left( 1 - T \right)$.
If the phase of $-T(s)$ at $f_{\left| T \right| = 1}$ becomes too close to 180$^{\circ}$, the overall loop gain turns positive and instability occurs.
The difference between $\pi$ and the phase of $-T(s)$ at $f_{\left| T \right| = 1}$ gives the phase margin, that is
\begin{equation}
\Phi_{PM}  = \pi - \frac{\pi}{2} - 2 \pi f_{\left| T \right| = 1} \tau_D = \frac{\pi}{2} -  \frac{\tau_D}{\tau_L}
\label{eq:PhiT1}
\end{equation}
In case of short signal and feedback lines the last term in (\ref{eq:PhiT1}) is negligible.
The phase margin is close to 90$^{\circ}$ and stability is assured.
If the signal and feedback lines are long, then the additional phase shift due to the propagation delay on both lines can affect stability.
Assuming the distance $L$ between the first and second stage to be fixed, the condition for a phase margin $\Phi_{PM} $ larger than 60$^{\circ}$ leads to a lower value for $\tau_L$, that is
\begin{equation}
\tau_L > \frac{6 \tau_D}{\pi} \simeq 2 \tau_D
\label{eq:tauL3}
\end{equation}
The loop gain is then stable provided that the condition (\ref{eq:tauL3}) is satisfied.
This can be achieved by properly choosing the value of $\tau_L$ by tuning the values of $C_P$ and $R_Q$ according to equation (\ref{eq:tauL1}) or (\ref{eq:tauL2}).

The evaluation of the phase shift introduced by the propagation delay along the cables is correct as long as both the signal and feedback lines are terminated at least at one end, in order to avoid multiple reflections.
In the schematic of figure \ref{fig:GeFROSchematic} both lines are terminated at the second stage, which seems the most convenient thing to do.
If either one of the lines were not terminated, the reflections back and forth would bear a larger phase shift than that given by the exponential term in $T(s)$, and the phase margin would be reduced, resulting in a more stringent lower limit for $\tau_L$.

\section{The ``Fast'' Signal}

Let us now derive the ``fast'' signal shape, as expressed by equation (\ref{eq:V1(t)1}).
Neglecting the feedback, and by approximating for frequencies above $1/ 2 \pi C_I R_F$, the open loop signal at the ``fast'' output in response to an instantaneous current pulse carrying a charge $Q$ in the complex frequency domain is given by
\begin{equation}
V_{1}^{OL} (s) = - G_1 \frac{Q}{s C_I} g_m R_T 
\label{eq:V1OL(s)1}
\end{equation}
where, as already discussed, $C_I$ is the total input capacitance and $g_m$ is the transconductance of $Q_1$.
Equation (\ref{eq:V1OL(s)1}) was obtained by approximating for $R_B \gg R_T$, and by considering the drain of the JFET as an ideal current source.
In the real case the output impedance of the JFET, contributed by its drain-source resistance and gate-drain capacitance, should be included in the calculations.
Its effect on the overall gain is a second order contribution and will not be considered here.
Equation (\ref{eq:V1OL(s)1}) also neglects the propagation delay due to the signal line length, which is a simple time shift of $\tau_D/2$.

As well known from feedback theory, the open loop signal $V_{1}^{OL}$ expressed by equation (\ref{eq:V1OL(s)1}) is modified by the presence of the feedback loop according to the relation
\begin{equation}
V_{1}(s) = \frac{V^{OL}_1 (s)}{1 - T(s)}
\label{eq:V1(s)2}
\end{equation}
Since the feedback loop is ineffective for frequencies above $1 / 2 \pi \tau_L$, and $\tau_L > \tau_D$ for stability, the denominator of (\ref{eq:V1(s)2}) can be approximated as
\begin{equation}
\frac{1}{1 - T (s)} = \frac{s \tau_L}{s \tau_L + \textrm{e}^{-s \tau_D}}
\simeq \frac{s \tau_L}{1+ s \left( \tau_L - \tau_D \right)} \frac{\tau_L - \tau_D}{\tau_L}
\end{equation}
The first term derives from a first order expansion of the exponential at frequencies smaller than $1 / 2 \pi \tau_D$.
The second term was introduced in order to satisfy the condition
\begin{equation}
\lim_{s \rightarrow \infty} \frac{1}{1 - T (s)} = 1 
\end{equation}
that is required since the feedback loop is ineffective at frequencies larger than $1/2 \pi \tau_L$.
Equation (\ref{eq:V1(s)2}) then becomes
\begin{equation}
V_{1}(s) = - G_1 \frac{Q}{s C_I} g_m R_T \frac{s \left(\tau_L - \tau_D \right)}{1 + s \left(\tau_L - \tau_D \right)}
\label{eq:V1(s)3}
\end{equation}
By taking the inverse Laplace transform of the above, one obtains the ``fast'' output signal expressed by equation (\ref{eq:V1(t)1}).
This first order approximation loses accuracy when $\tau_L$ is small and comparable to $\tau_D$.
In this case a second order expansion improves the accuracy, as shown in the appendix.

\section{The ``Slow'' Signal}

Let us now derive equation (\ref{eq:V2(t)1}), which gives the shape of the feedback signal or ``slow'' output.
If we consider the loop gain to be infinite, the input node is held at virtual ground and all the charge flows into $C_F$.
The signal in this case would then be given by
\begin{equation}
V_2^{IL} (s)  = - Q \frac{R_F}{1+s C_F R_F}
\label{eq:V2IL(s)}
\end{equation}
Again, the time shift due to the propagation delay along the cable was neglected.
Its effect on bandwidth is considered through the feedback loop gain $T(s)$.
Since the feedback loop has a finite gain and bandwidth, the actual closed loop signal differs from (\ref{eq:V2IL(s)}).
As well known from feedback theory it can be calculated as
\begin{equation}
V_2 (s)  = V_2^{IL} (s) \frac{- T(s)}{1 - T(s)}
\label{eq:V2(s)1}
\end{equation}
By using equation (\ref{eq:T(s)4}) for $T(s)$ we have that
\begin{equation}
 \frac{- T(s)}{1 - T(s)} = \frac{\textrm{e}^{-s \tau_D}}{\textrm{e}^{-s \tau_D} + s \tau_L}
\simeq \frac{1}{1+s\left( \tau_L - \tau_D \right)} 
\end{equation}
where again the exponential was approximated at first order for frequencies below $1 / 2 \pi \tau_D$, but a term $- s \tau_D$ at the numerator was dropped in order to satisfy the condition
\begin{equation}
\lim_{s \rightarrow \infty} \frac{-T(s)}{1 - T (s)} = 0 
\end{equation}
that is required since the ``slow'' signal does not contain high frequencies above the bandwidth of the feedback loop.
Equation (\ref{eq:V2(s)1}) then becomes
\begin{equation}
V_2 (s) = - \frac{Q}{s C_F} \frac{s \tau_F}{1 + s \tau_F} \frac{1}{1 + s \left(\tau_L-\tau_D  \right)}
\label{eq:V2(s)3}
\end{equation}
where $\tau_F = C_F R_F$.
By taking the inverse Laplace transform we obtain the ``slow'' output signal as expressed by equation (\ref{eq:V2(t)1}).
Again, a second order approximation can be considered to improve the accuracy for $\tau_L$ close to $\tau_D$, and is presented in the appendix.

\section{Signals at the Oscilloscope}
\label{sec:SigOsc}

\begin{figure}
\centering
\includegraphics[width=0.9\linewidth]{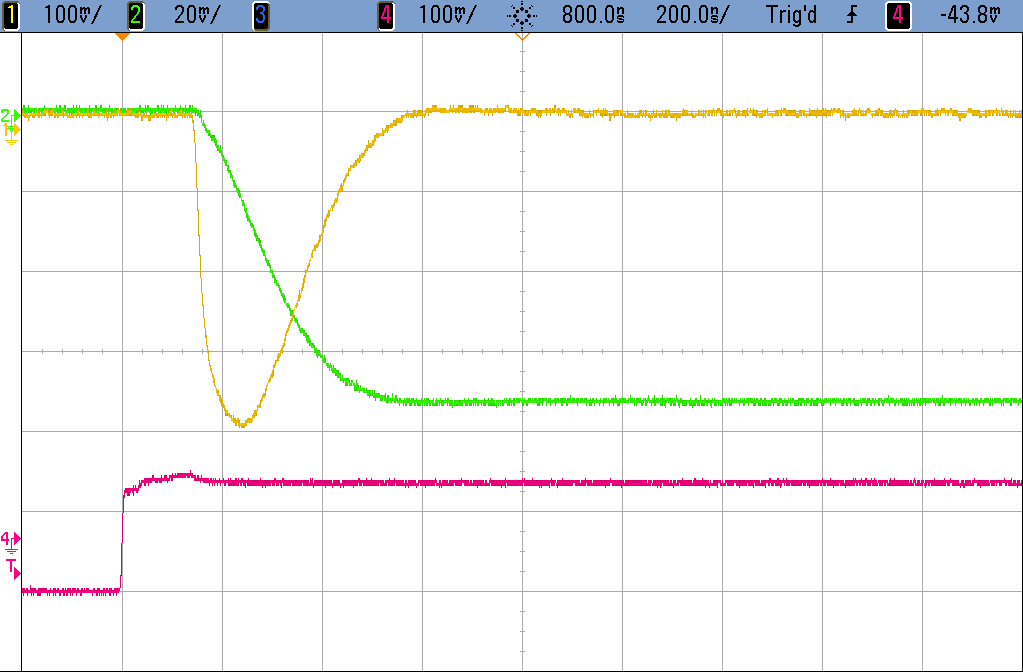}
\caption{Signals from the GeFRO in response to test charge pulses. The horizontal scale is \mbox{200 ns/div}. The pulser signal is shown in red (\mbox{100 mV/div}). The ``fast'' signal is shown in yellow (\mbox{100 mV/div}), the ``slow'' signal is shown in green (\mbox{20 mV/div}).}
\label{fig:GeFROScopePulser}
\end{figure}

As a first test of the validity of the above evaluations, the circuit was operated with the smallest possible value for $\tau_L$.
The input capacitance was \mbox{16 pF} due mainly to the input JFET.
The length of the cables between the first and second stage was \mbox{$L=$ 10 m}, so \mbox{$\tau_D =$ 100 ns}.
In this measurement the values \mbox{$R_P = \infty$}, \mbox{$C_P =$ 470 pF}, \mbox{$R_P =$ 33 $\Omega$} were chosen.
The gain-bandwidth product of the AD797 chosen for $A_2$ is \mbox{$\omega_T/ 2 \pi \simeq$ 100 MHz}.
With such choices \mbox{$1 / \omega_T R_Q \simeq$ 48 pF}, which makes its contribution negligible with respect to $C_P$ in (\ref{eq:tauL2}).
It is clear that with these values $R_P \gg R_Q$, as used in the above calculations, and \mbox{$C_P R_Q=$ 15 ns}, which makes the frequency of the zero fall outside the bandwidth of the feedback loop.
The value of $\tau_L$ which results from (\ref{eq:tauL2}) is  \mbox{$\tau_L = $ 184 ns}, which corresponds to a phase margin close to 60$^{\circ}$.
The expected 90\% to 10\% fall time of the ``fast'' signal (equal to the 10\% to 90\% rise of the ``slow'' signal) is  \mbox{2.2 $\left( \tau_L - \tau_D \right) =$ 185 ns}.

Figure \ref{fig:GeFROScopePulser} shows the outputs of the GeFRO as seen at the oscilloscope.
The image also shows the pulser signal used to simulate a charge pulse of \mbox{460 e$^{-}$}.
The rise time of the ``fast'' output is limited to a few tens of ns mainly by the bandwidth of the amplifier $A_1$.
The fall time of the ``fast'' output is about \mbox{180 ns}, and clearly coincides with the rise time of the ``slow'' output.
In figure \ref{fig:GeFROScopePulser} the fall of the ``slow'' signal cannot be seen in this time scale, since the 90\% to 10\% fall time \mbox{2.2 $C_F R_F$} is close to \mbox{1 ms}.
From the peak amplitude of the ``fast'' signal, knowing the values of $G_1$, $Q$, $g_m$ and $R_T$, the input capacitance $C_I$ can be measured with the oscilloscope.
As can be seen in figure \ref{fig:GeFROScopePulser}, the amplitude of the ``fast'' output in response to a test pulse of \mbox{460 ke$^{-}$} was \mbox{390 mV}.
By adding at the input a known \mbox{10 pF} capacitor to ground, and adjusting $R_Q$ by hand to obtain the same value for $\tau_L$, the amplitude of the ``fast'' signal decreased to about \mbox{240 mV}.
From this it is possible to estimate the input capacitance in the previous case as \mbox{$C_{J} =$ 16 pF}, mainly given by the JFET.

\begin{figure}
\centering
 \includegraphics[width=0.9\linewidth]{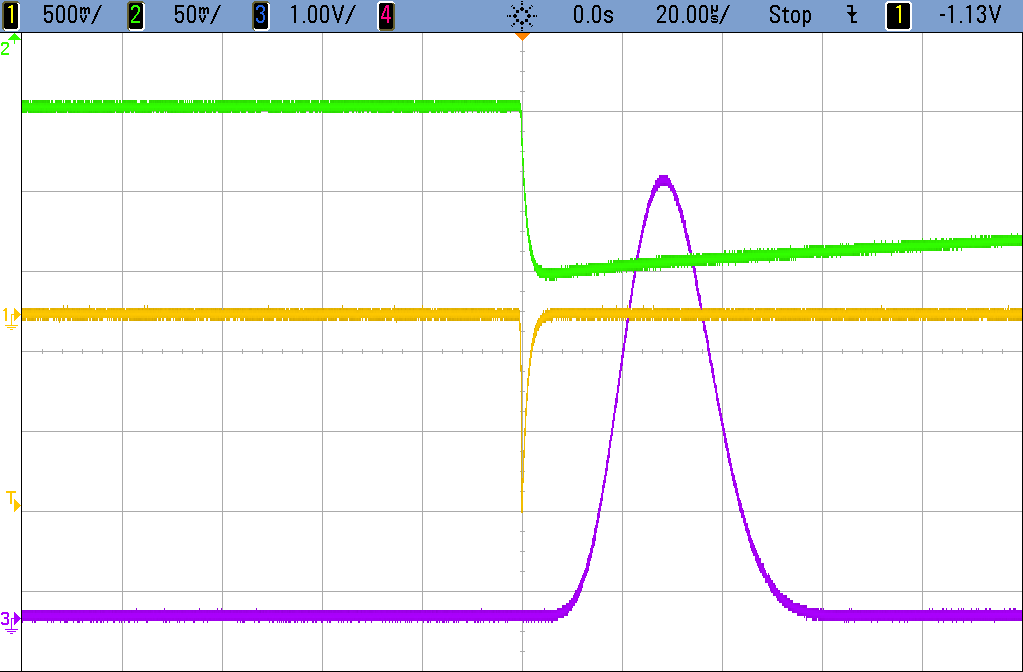}
\phantom{.}
 \includegraphics[width=0.9\linewidth]{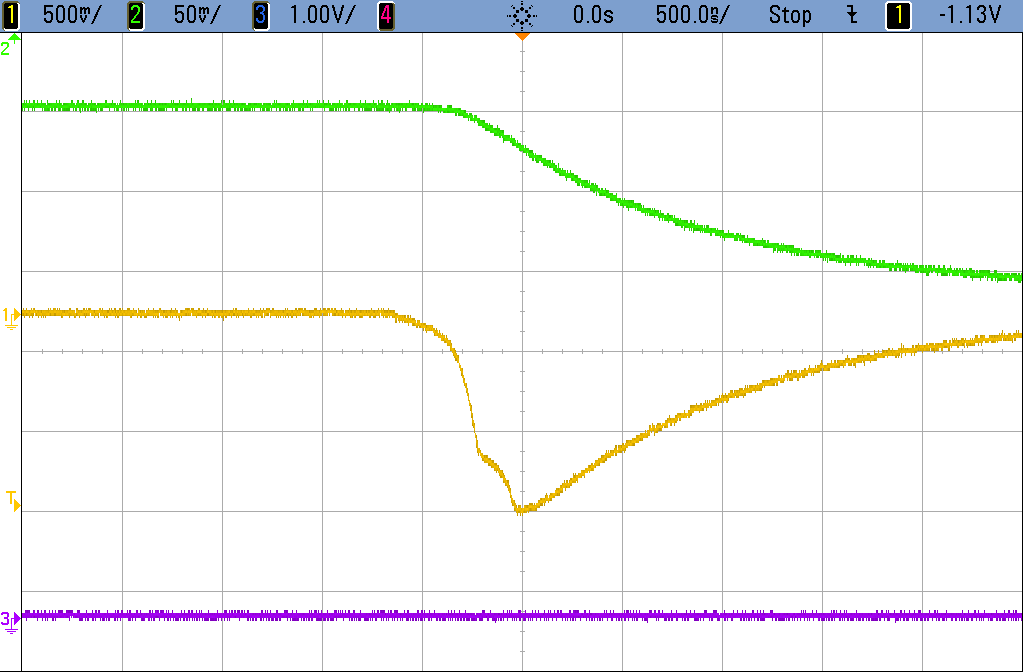} 
\caption{Signals from the GeFRO coupled to a Canberra BEGe detector. The ``fast'' signal is shown in yellow (\mbox{500 mV/div}), the ``slow'' signal is shown in green (\mbox{50 mV/div}).
The shaped Gaussian signal used to acquire the energy spectrum is shown in purple.
The horizontal scale is \mbox{20 $\mu$s/div} at the top, \mbox{500 ns/div} at the bottom.}
\label{fig:GeFROScopeSignal}
\end{figure}

The known \mbox{10 pF} capacitor was then disconnected, and a Canberra BEGe detector was connected at the input of the circuit with a short wire.
In these conditions the amplitude of the ``fast'' signal was again close to \mbox{240 mV}.
From this it is possible to infer that the capacitance added by the BEGe detector (and connection parasitics) is \mbox{$C_D=$ 10 pF}.
The total capacitance at the input with the GeFRO circuit coupled to the detector was then \mbox{$C_I = C_J + C_D =$ 26 pF}, that is the value which was already considered in the previous calculations.
The values of $R_P$ and $R_S$ were then changed to \mbox{$R_P=$ 820 k$\Omega$} and \mbox{$R_Q=$ 100 $\Omega$} in order to obtain the optimal working conditions with the BEGe detector.
These values satisfy the condition (\ref{eq:CFRF}) and result in \mbox{$\tau_L \simeq $ 950 ns}, as calculated from (\ref{eq:tauL1}).
By acting on the amplifier $A_1$, the gain of the ``fast'' signal $G_1$ was then doubled with respect to the previous case.

Figure \ref{fig:GeFROScopeSignal} shows the signals seen with the oscilloscope when the detector was illuminated with a $^{228}$Th gamma source.
The figure shows the ``fast'' and ``slow'' signals in yellow and green respectively for a given event of energy close to \mbox{2 MeV}.
The upper image in figure  \ref{fig:GeFROScopeSignal} was taken with a time scale of \mbox{20 $\mu$s/div}.
The lower image shows the same event on a time scale of \mbox{500 ns/div}.
The 90\% to 10\% fall time of the ``fast'' signal is about \mbox{1.8 $\mu$s} with the values chosen above for $C_P$, $R_P$ and $R_Q$.
A larger value for the fall time was chosen with respect to the previous case, since as reported in \cite{psa1, psa2} the charge collection time in BEGe detectors is relatively slow, ranging up to a few hundred nanoseconds.
As can be seen in the lower image, the high timing resolution of the ``fast'' signal faithfully reproduces the charge collection profile in the BEGe detector.
The event in the figure is clearly a multi-site event, showing separate steps in the charge collection profile.
Figure \ref{fig:GeFROScopeSignal} also shows the ``slow'' signal after Gaussian shaping at \mbox{10 $\mu$s}, obtained with an Ortec 672 shaper, which was used to measure the energy spectra with an Ortec 919 multichannel analyzer.
A more detailed discussion of noise and energy resolution will be given in the following sections.

\section{Noise}

The ``slow'' output of the GeFRO can be shaped with proper filters and used for energy measurements.
The most common case in analog processing is Gaussian shaping, already shown in figure \ref{fig:GeFROScopeSignal}.
The amplitude of the Gaussian signal is proportional to the deposited charge, which is in turn proportional to the total energy deposited in the detector by a particle event.
The RMS noise of the circuit adds in quadrature to the intrinsic resolution of the detector, and the result gives the expected energy resolution of the system.
The noise at the ``slow'' output of the GeFRO circuit after a Gaussian shaper with time constant $\tau$ can be evaluated from the well known equivalent noise charge formula
\begin{equation}
\sigma_Q = \sqrt{   i_n^2 \beta \tau  +  A_f C_I^2 \gamma +  e_n^2 C_I^2 \frac{\alpha}{ \tau}   }
\label{eq:sigmaQ}
\end{equation}
where $i_n$ is the white current noise spectral density, $C_I$ is the total input capacitance, $A_f$ is the 1/f voltage noise coefficient and $e_n$ is the white voltage noise spectral density.
In the case of Gaussian shaping the coefficients $\alpha$, $\beta$ and $\gamma$ take the values
\mbox{$\alpha \simeq$ 0.44}, \mbox{$\beta \simeq$ 0.89}, \mbox{$\gamma \simeq$ 3.14}.

\begin{figure}
\centering
\includegraphics[scale=0.45]{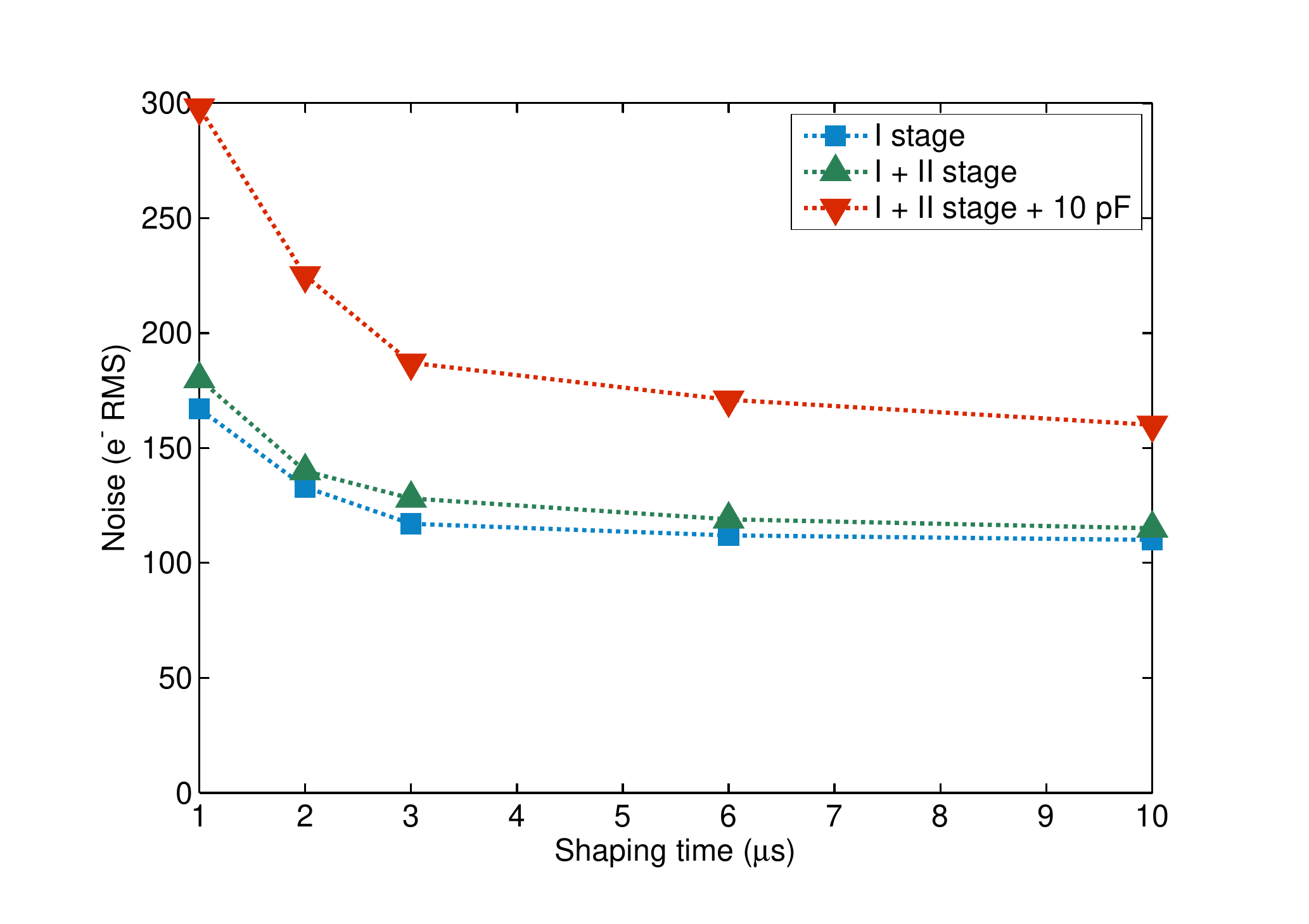}
\caption{Equivalent noise charge from the first stage alone (blue curve), from the first and second stage (green curve), and from the first and second stage after the addition of a \mbox{10 pF} capacitor to ground at the input to simulate the detector (red curve).}
\label{fig:GeFRONoiseVSTau}
\end{figure}

\begin{figure*}
\centering
\includegraphics[scale=0.45]{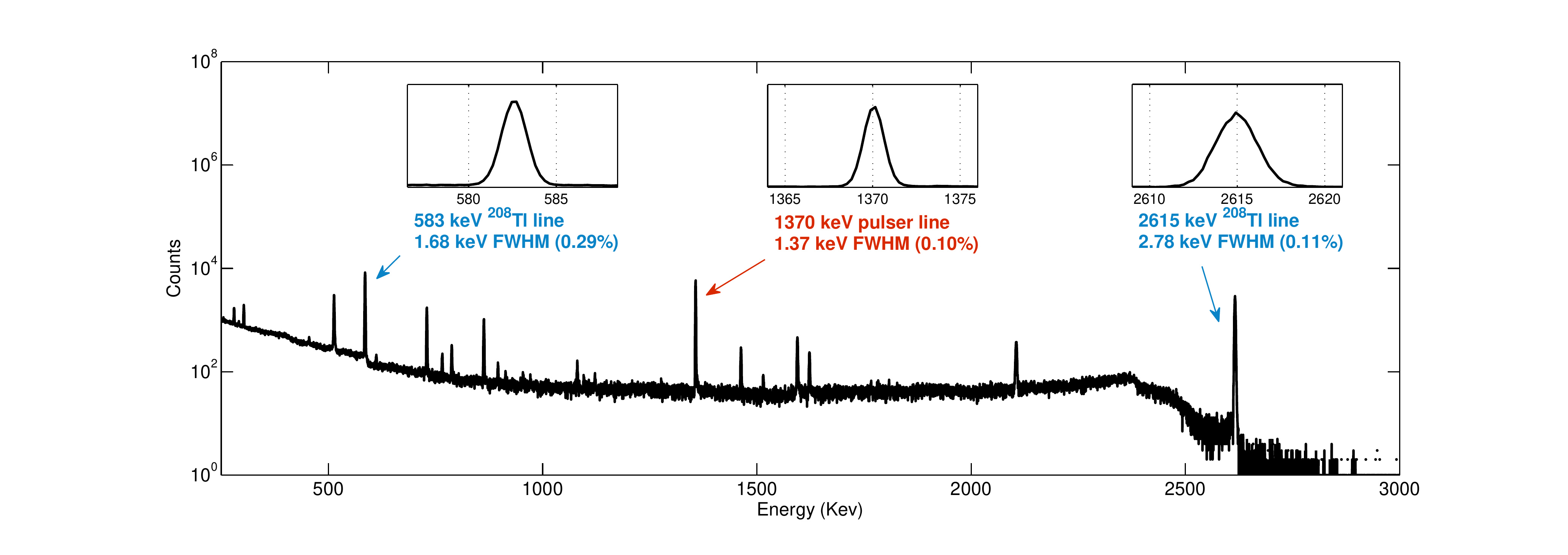}
\caption{$^{228}$Th spectrum obtained with a BEGe detector readout with the GeFRO.}
\label{fig:GeFROSpectrum}
\end{figure*}

Let us first consider the current noise sources.
At \mbox{10 $\mu$s} shaping a detector leakage current of \mbox{1 pA} gives a shot noise of \mbox{0.6 fA/$\sqrt{\textrm{Hz}}$} and contributes to $\sigma_Q$ with about \mbox{10 e$^-$ RMS}.
The feedback resistor $R_F$, whose value is \mbox{500 M$\Omega$} and is held at \mbox{77 K}, gives a thermal noise of \mbox{2.9 fA/$\sqrt{\textrm{Hz}}$}, and thus contributes with \mbox{55 e$^-$ RMS}.
The total current noise at the input at \mbox{10 $\mu$s} is then just below \mbox{60 e$^-$ RMS}, dominated by $R_F$.
As expressed by (\ref{eq:sigmaQ}) the weight of the current noise decreases at shorter shaping times.

If the gain of the cold stage were much larger than one, the noise sources at the second stage could be neglected, and all the series noise would be given by the input transistor $Q_1$.
This was directly measured by terminating the signal line with a resistor $R_T =$ 200 $\Omega$, obtaining a gain of 6.6 between the first and second stage.
The value of $\tau_L$ was adjusted to a few hundred ns, allowing to shape the ``slow'' signal with the gaussian shaper with time constant from \mbox{1 $\mu$s} to \mbox{10 $\mu$s}.
The equivalent noise charge was evaluated by dividing the RMS noise at the output, measured with a Rohde\&Schwartz URE3 RMS voltmeter, by the peak amplitude of the Gaussian shaped signal in response to a known test pulse.
The resulting values for the noise of the first stage with \mbox{$C_I=$ 16 pF} are shown in figure \ref{fig:GeFRONoiseVSTau}, blue curve.
From this measurement, the series white noise of the input transistor can be evaluated at \mbox{1 $\mu$s}, obtaining a white noise density of \mbox{2.5 nV/$\sqrt{\textrm{Hz}}$}.
The noise at \mbox{10 $\mu$s} is instead dominated by the 1/f contribution, together with the current noise from $R_F$.
After substracting the latter we are left with about \mbox{100 e$^-$ RMS}, that is compatible with a value for the 1/f noise coefficient of \mbox{$A_f=$ 3$\cdot$10$^{-13}$ V$^2$}.

When the gain of the first stage is brought back to the original value with $R_T =$ 50 $\Omega$, the noise from the second stage must be considered.
In our case \mbox{$g_m R_T\simeq$ 1.7}, so the noise from the termination resistor $R_T$, of the resistor $R_Q$ and of the second stage amplifier should be considered at the input divided by \mbox{$g_m R_T$} and summed in quadrature to the noise contribution of the JFET in the evaluation of the overall series noise.
The resistor $R_Q$ contributes with \mbox{0.8 nV/$\sqrt{\textrm{Hz}}$} at the input, while $R_T$ contributes with \mbox{0.5 nV/$\sqrt{\textrm{Hz}}$}. The white voltage noise from $A_2$ referred to the input of $Q_1$ is about \mbox{0.5 nV/$\sqrt{\textrm{Hz}}$}.
The current noise of $A_1$ and $A_2$, about \mbox{2 pA/$\sqrt{\textrm{Hz}}$} in both cases, would also add to the series noise, but their contribution is negligible since the impedance $R_T$ seen at their inputs is small.
The total white voltage noise at the input is then close to \mbox{1 nV/$\sqrt{\textrm{Hz}}$}, which at \mbox{10 $\mu$s} with an input capacitance of \mbox{16 pF} gives about \mbox{20 e$^-$ RMS}.
With a proper choice of the amplifier $A_2$, its 1/f voltage and current noise can be neglected.
The same is true for the current noise contribution from $A_1$.
Even if the gain of the first stage is only 1.7, the contribution of the second stage is then small compared to that of the first stage, as can be clearly seen from the green curve of figure \ref{fig:GeFRONoiseVSTau}, which is only slightly above the blue curve.

Finally, a \mbox{10 pF} capacitor was added at the input to simulate the detector, obtaining the red curve in figure \ref{fig:GeFRONoiseVSTau}.
From the comparison of this curve with the others in the same figure, it is clearly evident that with a total input capacitance of \mbox{26 pF} at \mbox{10 $\mu$s} shaping time the equivalent noise charge of the shaped ``slow'' signal is dominated by the 1/f series noise of the input JFET, and is close to \mbox{160 e$^{-}$ RMS}.


\section{Energy spectra}
\label{sec:EneSpe}

\begin{figure}
\centering
\includegraphics[scale=0.45]{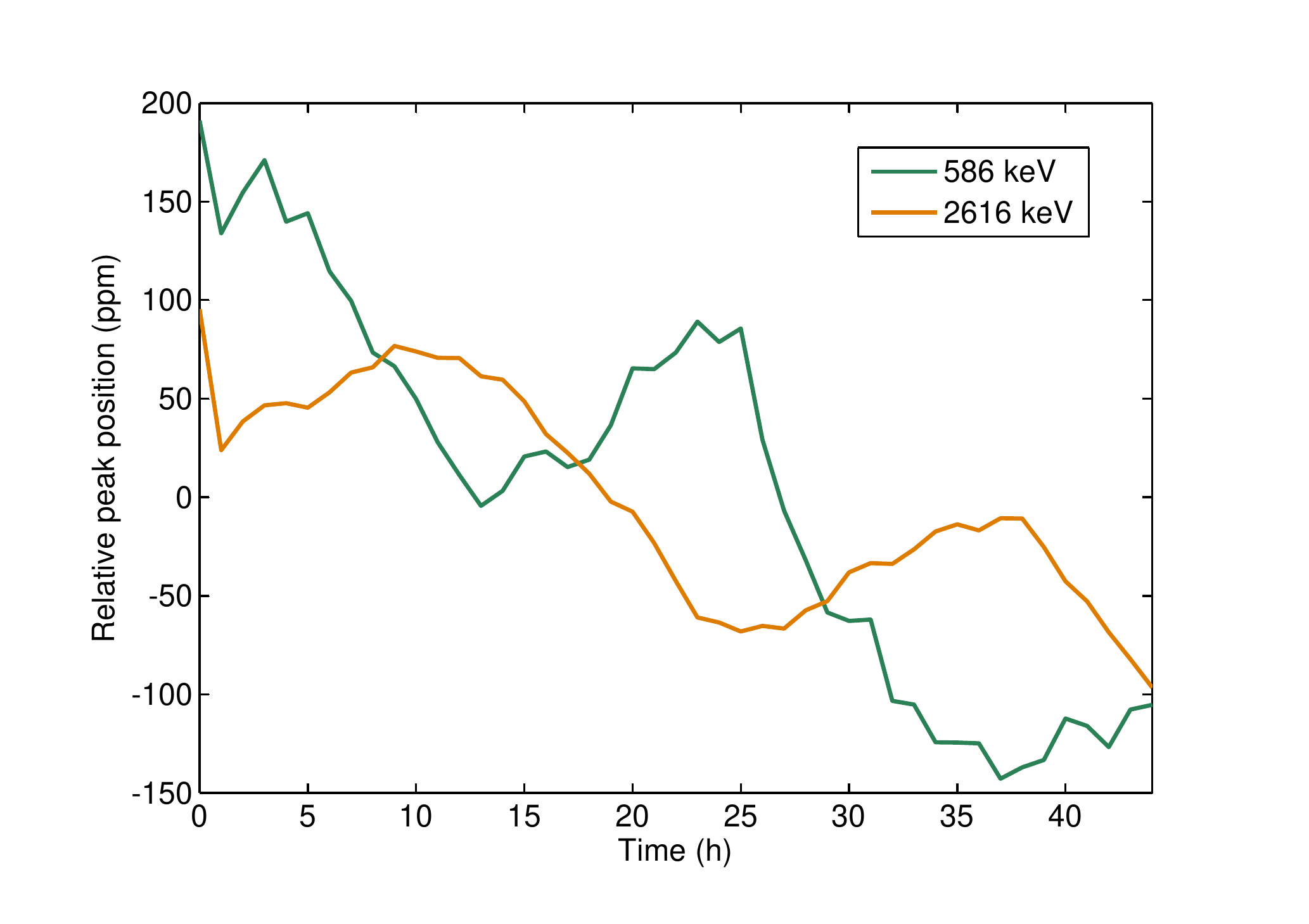}
\caption{Relative position of the \mbox{583 keV} and \mbox{2615 keV} peaks over 44 hours of measurement.}
\label{fig:GeFRODrift}
\end{figure}

The equivalent noise charge of \mbox{160 e$^{-}$ RMS}, as results from the previous section, corresponds in Germanium to a FWHM resolution of about \mbox{1.1 keV FWHM}.
This can be verified in the spectrum shown in figure \ref{fig:GeFROSpectrum}, taken by facing the BEGe detector with a $^{228}$Th gamma source.
The resolution of the pulser line set at \mbox{1370 keV} is \mbox{1.37 keV FWHM}, higher than expected, most likely due to some small disturbance injected through the high voltage power supply for the detector.
The result is anyway remarkable.
The resolution of the \mbox{583 keV} and \mbox{2615 keV} lines is \mbox{1.68 keV} and \mbox{2.78 keV} respectively, as expected from the Poisson statistics compensated by the Fano factor in Germanium, summed in quadrature with the electronic noise.
The spectrum was taken over an hour with an overall event rate of about 800 counts per second.

The drift in the position of the \mbox{583 keV} and \mbox{2615 keV} lines versus time is shown in figure \ref{fig:GeFRODrift}.
The figure shows a continuous drift for both lines close to \mbox{4 ppm/h}, likely related with the evaporation of liquid Nitrogen in the dewar housing the detector, which was not refilled in 44 hours, and a daily periodic trend, related with small temperature and humidity variations at the second stage.
Curiously the daily variations for the two lines are not in phase.
In any case, the measurement shows a very good overall stability in the position of the peaks, better than about \mbox{10 ppm/h} over 44 hours of measurement.

\section{Conclusions}

The readout chain for semiconductor detectors presented in this paper provides a novel way to solve the trade-off between wide bandwidth, high energy resolution and the requirement of a minimal number of front-end components close to the detector.
By acquiring both the ``fast'' and the ``slow'' outputs of the GeFRO all the relevant information from the detector signals can be retrieved.
The design approach of the GeFRO circuit was described in detail, together with the criteria for component selection and the trade-offs involved.

The circuit was tested with a Canberra BEGe detector demonstrating a high timing resolution at the ``fast'' output, enough to clearly resolve the charge collection profiles in the detector for single-site and multi-site events.
At the same time the spectra measured from the shaped ``slow'' output provided high resolution and very reliable operation, with a negligible drift in peak position over several hours.
These features make the GeFRO circuit particularly suitable for rare event searches with semiconductor detectors, and in particular for the neutrinoless double beta decay search experiments GERDA and MAJORANA.

\appendix

Here the calculations for the shape of the ``fast'' output will be carried out to a second order approximation in $\tau_D/\tau_L$.
By approximating the loop gain (\ref{eq:T(s)4}) at the second order in $\tau_D/\tau_L$, and without introducing normalization factors, we obtain
\begin{equation}
T (s) \simeq - \frac{1 - s \tau_D + s^2 \tau_D^2 /2}{s \tau_L}
\label{eq:T(s)5}
\end{equation}
By plugging this into (\ref{eq:V1(s)2}), the ``fast'' signal takes the form
\begin{equation}
V_{1}(s) = - G_1 \frac{Q}{s C_I} g_m R_T \frac{ s \tau_L}{1 + s \left(\tau_L - \tau_D  \right) + s^2 \tau_D^2 /2}
\end{equation}
Let us now consider the limit case which satisfies (\ref{eq:tauL3}), that is $\tau_L = 2 \tau_D$.
The above equation becomes
\begin{equation}
V_{1}(s) = - G_1 \frac{Q}{s C_I} g_m R_T \frac{ 2 s \tau_D}{1 + s \tau_D + s^2 \tau_D^2 /2}
\end{equation}
The inverse Laplace transform gives
\begin{equation}
V_{1}(t) = - G_1 \frac{Q}{C_I} g_m R_T \left(4 \textrm{e}^{ -\frac{t}{\tau_D}} \sin{\frac{t}{\tau_D}}\right)
\label{eq:V1(t)2}
\end{equation}
The signal expressed by (\ref{eq:V1(t)2}) reaches its maximum value at $t_{M} = \tau_D {\pi}/{4}$, that is about \mbox{80 ns} if \mbox{$\tau_D =$ 100 ns}.
Its amplitude at $t_{M}$ is
\begin{equation}
V_1 \left( t_{M} \right) = 1.29\ G_1 \frac{Q}{C_I} g_m R_T
\end{equation}
Its 10\% to 90\% rise time is \mbox{46 ns}, its 90\% to 10\% fall time is \mbox{152 ns}.
The signal in (\ref{eq:V1(t)2}) has a small undershoot below ground, less than 5\% of the total amplitude.
Concerning the ``slow'' signal, from (\ref{eq:V2(s)1}) and (\ref{eq:T(s)5}) we find
\begin{equation}
V_2 (s) = - Q \frac{Q}{s C_F} \frac{s \tau_F}{1 + s \tau_F} \frac{1 - s \tau_D + s^2 \tau_D^2 /2}{1 + s \left(\tau_L-\tau_D  \right) + s^2 \tau_D^2 /2}
\end{equation}
Again, the expression can be evaluated for the limit case \mbox{$\tau_L = 2 \tau_D$}, obtaining
\begin{equation}
V_2 (s) = - Q \frac{Q}{s C_F} \frac{s \tau_F}{1 + s \tau_F} \frac{1 - s \tau_D + s^2 \tau_D^2 /2}{1 + s \tau_D + s^2 \tau_D^2 /2}
\end{equation}
and by calculating the inverse Laplace transform one obtains
\begin{equation}
V_{2}(t) = - \frac{Q}{C_F} \left( \textrm{e}^{ -\frac{t}{\tau_F}} -  4 \textrm{e}^{ -\frac{t}{\tau_D}} \sin{\frac{t}{\tau_D}} \right)
\label{eq:V2(t)2}
\end{equation}
Its amplitude is ${Q}/{C_F}$, its 10\% to 90\% rise time is \mbox{152 ns} and its 90\% to 10\% fall time is \mbox{2.2 $\tau_F$}.
Some of the features of the ``fast'' and ``slow'' signals as given by equations (\ref{eq:V1(t)2}) and (\ref{eq:V2(t)2}) can be directly seen in figure \ref{fig:GeFROScopePulser}, which was taken for \mbox{$\tau_L \simeq$ 2$\tau_D$}.


\begin{thebibliography}{3}

\bibitem{chargeamp1}
E. Fairstein,
\textit{Considerations in the Design of Pulse Amplifiers for Use with Solid State Radiation Detectors},
IRE Transactions on Nuclear Science {8} (1961) 129.\\
doi: 10.1109/TNS2.1961.4315811

\bibitem{chargeamp2}
V. Radeka,
\textit{The field-effect transistor-its characteristics and applications},
IEEE Transactions on Nuclear Science {11} (1964) 358.\\
doi: 10.1109/TNS.1964.4323448

\bibitem{chargeamp3}
T. V. Blalock,
\textit{A Low-Noise Charge-Sensitive Preamplifier with a Field-Effect Transistor in the Input Stage},
IEEE Transactions on Nuclear Science {11} (1964) 365.\\
doi: 10.1109/TNS.1964.4323449

\bibitem{gefro1}
C. Cattadori, B. Gallese, A. Giachero, C. Gotti, M. Maino, G. Pessina,
\textit{A new approach to the readout of cryogenic ionization detectors: GeFRO},
Journal of Instrumentation {6} P05006 (2011).\\
doi: 10.1088/1748-0221/6/05/P05006

\bibitem{gefro2}
C. Cattadori, A. Giachero, C. Gotti, M. Maino, G. Pessina,
\textit{GeFRO, a new front-end approach for the phase II of the GERDA experiment},
IEEE Nuclear Science Symposium Conference Record (2011) 1463.\\
doi: 10.1109/NSSMIC.2011.6154349

\bibitem{gerda1}
The GERDA Collaboration,
\textit{The GERDA experiment for the search of 0nbb decay in 76Ge},
European Physical Journal C {73} (2013) 2330.\\
doi: 10.1140/epjc/s10052-013-2330-0

\bibitem{majorana1}
The MAJORANA Collaboration,
\textit{The MAJORANA experiment: an ultra-low background search for neutrinoless double-beta decay},
Journal of Physics: Conference Series {381} (2012) 012044.\\
doi: 10.1088/1742-6596/381/1/012044

\bibitem{psa1}
D. Budjas, M. Barnabe-Heider, O. Chkvorets, N. Khanbekov, S. Schonert,
\textit{Pulse shape discrimination studies with a Broad-Energy Germanium detector for signal identification and background suppression in the GERDA double beta decay experiment},
Journal of Instrumentation {4} (2009) P10007.\\
doi: 10.1088/1748-0221/4/10/P10007

\bibitem{psa2}
R. J. Cooper, D. C. Radford, K. Lagergren, J. F. Colaresi, L. Darken, R. Henning, M. G. Marino, K. Michael Yocum,
\textit{A pulse shape analysis technique for the MAJORANA experiment},
Nuclear Instruments and Methods A {629} (2011) 303.\\
doi: 10.1016/j.nima.2010.11.029

\bibitem{psa3}
M. Agostini, C. A. Ur, D. Budjas, E. Bellotti, R. Brugnera, C. M. Cattadori, A. di Vacri, A. Garfagnini, L. Pandola, S. Schönert,
\textit{Signal modeling of high-purity Ge detectors with a small read-out electrode and application to neutrinoless double beta decay search in Ge-76},
Journal of Instrumentation {6} P03005 (2011).\\
10.1088/1748-0221/6/03/P03005

\end{thebibliography}
\end{document}